\documentstyle[aps,prl,epsfig]{revtex}

\bibstyle{unsrt}
\begin{document}
\draft

\title{Two-Loop Euler-Heisenberg QED Pair-Production Rate}
\author{Gerald V. Dunne$^1$ and Christian Schubert$^2$}
\address{${}^1$Department of Physics, University of Connecticut, Storrs CT
06269, USA}
\address{${}^2$Laboratoire d'Annecy-le-Vieux de
Physique Th{\'e}orique LAPTH, F-74941 Annecy-le-Vieux, France}

\maketitle
\vskip .5cm

\begin{abstract}
We study the divergence of large-order perturbation theory in the
worldline expression for the two-loop Euler-Heisenberg QED effective
Lagrangian in a constant magnetic field. The leading rate of divergence is
identical, up to an overall factor, to that of the one-loop case. From
this we deduce, using Borel summation techniques, that the leading behaviour  
of the imaginary part of the two-loop effective Lagrangian for a constant E
field, giving the pair-production rate, is proportional to the one-loop
result. This also serves as a test of the mass renormalization, and
confirms the earlier analysis by Ritus.
\end{abstract}

\vskip 1cm
\hspace{12cm}LAPTH-739/99

\vskip 1cm

\section{Introduction}

Euler and Heisenberg \cite{euler}, and many others since
\cite{weisskopf,schwinger,nikishov,wd,greiner}, computed the exact renormalized  
one-loop QED effective action for electrons in a uniform electromagnetic field  
background. When the background is purely that of a static magnetic field,
the effective action is minus the effective energy of the electrons in that
background. When the background is purely that of a uniform electric field,
the effective action has an imaginary part which determines the
pair-production rate of electron-positron pairs from vacuum
\cite{euler,weisskopf,schwinger,nikishov,wd,greiner}.
In this paper we consider the two-loop Euler-Heisenberg effective action, and  
we show how the divergence of the perturbative expression for the effective
action with a uniform magnetic background is related to the non-perturbative  
imaginary part of the effective action with a uniform electric background.
The two-loop Euler-Heisenberg effective
Lagrangian, describing the effect of
a single photon exchange in the electron loop, was first calculated
by Ritus \cite{ritus}, and later recalculated by
Dittrich and Reuter \cite{dittrich} for the magnetic field case.
In both cases the proper-time
method and the exact Dirac propagator in the
uniform field were used.
More recently the magnetic field computation was
repeated in \cite{rss} using the more convenient `worldline' formalism  
\cite{strassler,cs}.
This calculation revealed that the
previous results by Ritus and Dittrich/Reuter were
actually incompatible, and differed precisely by a finite
electron mass renormalization. This prompted yet another recalculation of
this quantity in the worldline formalism \cite{frss}, now using
dimensional regularisation instead of a proper-time cutoff
as had been used in the previous calculations.
That calculation confirmed the correctness of Ritus's result,
and conversely showed that the final result given by \cite{dittrich}
was not expressed in terms of the physical electron mass. As part of
our analysis here, we show how this finite difference in the mass
renormalization affects the large-order behaviour of perturbation theory, and  
how this affects the leading contribution to the imaginary part of the
effective action in the electric field case.

For a uniform magnetic background, of strength $B$, the one-loop effective
Lagrangian has a simple ``proper-time'' integral representation
\cite{euler,weisskopf,schwinger,nikishov,wd,greiner}:
\begin{equation}
{\cal L}^{(1)}=-\frac{m^4}{8 \pi^2}\left(\frac{e B}{m^2}\right)^2
\int_{0}^{\infty} \frac{ds}{s^{2}}\;
(\coth s-\frac{1}{s}-\frac{s}{3})\,e^{-m^2s/(eB)}
\label{proper}
\end{equation}
The $\frac{1}{s}$ term is a subtraction of the zero field ($B=0$) effective
action, while the $\frac{s}{3}$ subtraction corresponds to a logarithmically  
divergent charge renormalization. For a given strength $B$, this integral can  
be evaluated numerically \cite{dittrich}. Alternatively, we can make contact with a
perturbative evaluation of the one-loop effective action by making an
asymptotic expansion of the integral in the weak field limit -- i.e., for
small values of the dimensionless parameter $\frac{eB}{m^2}$:
\begin{eqnarray}
{\cal L}^{(1)} \sim -\frac{2 m^4}{\pi^2}\left(\frac{e B}{m^2}\right)^4 \,
\sum_{n=0}^\infty
\frac{2^{2n} {\cal B}_{2n+4}} {(2n+4)(2n+3)(2n+2)}\left(\frac{eB}{m^2}\right)^{2n}
\label{eh}
\end{eqnarray}
Here the ${\cal B}_{2n}$ are Bernoulli numbers \cite{gradshteyn}. Each term
in this expansion of ${\cal L}^{(1)}$ is associated with a one-fermion-loop
Feynman diagram. Note that only even powers of $eB$ appear, as expected due to  
charge conjugation invariance (Furry's theorem). The divergent ${\rm O}(e^2)$  
self-energy term is not included as it contributes to the bare Lagrangian by  
charge renormalization.

The expansion (\ref{eh}) is the prototypical ``effective field theory''
effective Lagrangian \cite{donoghue}, where the low energy effective
Lagrangian, for energies well below the fermion mass scale $m$, is expanded as
\begin{equation}
{\cal L}=m^4\, \sum_n a_n\, {O^{(n)}\over m^n}
\label{eff}
\end{equation}
with $O^{(n)}$ being an operator of dimension $n$. For QED in a
uniform background, the higher dimensional operators $O^{(n)}$ are
formed from powers of Lorentz invariant combinations of the uniform
field strength $F_{\mu\nu}$. For a
uniform magnetic background this simply means even powers of $B$, as in
(\ref{eh}). Note that the `low energy' condition here means that the cyclotron  
energy $\frac{eB}{m}$ is well below the fermion mass scale $m$; in other
words, $\frac{eB}{m^2}\ll 1$.

The Euler-Heisenberg Lagrangian encodes the information on the
low-energy limit of the one-loop $N$ - photon amplitudes
in a way which is highly convenient for the derivation of
various nonlinear QED effects such as vacuum birefringence
(see, e. g., \cite{ditgie} and refs. therein) or
photon splitting \cite{biabia,adler71}.
The experimental observation of
vacuum birefringence is presently attempted by laser
experiments \cite{pvlas,fermilab877}.
There is also recent experimental evidence
for vacuum effects in pair production with strong laser electric fields
\cite{burkeetal}.

The one-loop Euler-Heisenberg perturbative effective action (\ref{eh}) is not  
a convergent series. The one-loop expansion coefficients in (\ref{eh})
alternate in sign [since ${\rm sign}({\cal B}_{2n})=(-1)^{n+1}$], but grow
factorially in magnitude (see also Table 1):
\begin{eqnarray}
a_n^{(1)}&=&-\frac{2^{2n}{\cal B}_{2n+4}} {(2n+4)(2n+3)(2n+2)}\nonumber\\
&\sim& (-1)^{n}\frac{1}{8 \pi^4}\,\frac{\Gamma(2n+2)}{\pi^{2n}}\left(
1+\frac{1}{2^{2n+4}}+ \frac{1}{3^{2n+4}}+\dots\right)
\label{growth}
\end{eqnarray}
So the perturbative expansion (\ref{eh}) is a divergent series. This
divergent behaviour is not a bad thing; it is completely analogous to generic  
behaviour that is well known in perturbation theory in both quantum field
theory and quantum mechanics \cite{zinn}. For example, Dyson \cite{dyson}
argued physically that QED perturbation theory is non-analytic, and
therefore presumably divergent,
as an expansion in the fine structure constant $\alpha$,
because the theory
is unstable when $\alpha$ is negative. As is well known, the divergence of
high orders of perturbation theory can be used to extract information about
non-perturbative decay and tunneling rates, thereby providing a bridge between  
perturbative and non-perturbative physics \cite{zinn}. It has been
argued \cite{zhitnitsky}, based on the behaviour of the one-loop
Euler-Heisenberg effective
Lagrangian (\ref{eh}), that the effective field theory expansion
(\ref{eff}) is generically divergent. Here we consider this question at the
two-loop level.

We stress that for energies well below the scale set by the fermion mass $m$,  
the divergent nature of the effective Lagrangian is not important, as the
first few terms in the series (\ref{eh}) provide an accurate approximation.
However, the divergence properties do become important when the external
energy scale approaches the fermion mass scale $m$. The divergence is also the  
key to understanding how non-perturbative {\it imaginary} contributions to
the effective action arise from {\it real} perturbation theory.

\section{Borel Analysis of the One-Loop Euler-Heisenberg Effective Lagrangian}

To begin, we review very briefly some basics of Borel summation
\cite{hardy,carl,zinnborel,thooft}. Consider an asymptotic series expansion  
of some function
$f(g)$
\begin{eqnarray}
f(g)\sim \sum_{n=0}^\infty \, a_n\, g^n
\label{exp}
\end{eqnarray}
where $g\to 0^+$ is a small dimensionless perturbation expansion parameter.
In an extremely broad range of physics applications \cite{zinn} one finds
that perturbation theory leads not to a convergent series but to a divergent  
series in which the expansion coefficients $a_n$ have leading large-order
behaviour
\begin{eqnarray}
a_n\sim (-1)^n \rho^n \Gamma(\mu\, n+\nu)  \qquad\qquad (n\to\infty)
\label{general}
\end{eqnarray}
for some real constants $\rho$, $\mu>0$, and $\nu$. When $\rho>0$, the
perturbative expansion coefficients $a_n$ alternate in sign and their
magnitude grows factorially, just as in the Euler-Heisenberg case
(\ref{growth}). Borel summation is a useful approach to this case of a
divergent, but alternating series. Non-alternating series must be treated
somewhat differently.

To motivate the Borel approach, consider the classic example :
$a_n=(-1)^n \rho^n n!$, and $\rho>0$. The series (\ref{exp}) is clearly
divergent for any value of the expansion parameter $g$. Write
\begin{eqnarray}
f(g)&\sim & \sum_{n=0}^\infty (-1)^n (\rho g)^n \, \int_0^\infty ds\, s^n\,
e^{-s}\nonumber\\
&\sim&
\frac{1}{\rho g}\, \int_0^\infty \,ds\, \left({1\over 1+s}\right) \,
\exp\left[- \frac{s}{\rho g}\right]
\label{borel}
\end{eqnarray}
where we have formally interchanged the order of summation and integration.
The final integral, which is convergent for all $g >0$, is {\it defined} to
be the sum of the divergent series. To be more precise \cite{hardy,carl}, the  
formula (\ref{borel}) should be read backwards: for $g\to 0^+$, we can use
Laplace's method to make an asymptotic expansion of the integral, and we
obtain the asymptotic series in (\ref{exp}) with expansion coefficients
$a_n=(-1)^n \rho^n n!$.

For a non-alternating series, such as $a_n=\rho^n n!$, we need $f(-g)$. The
Borel integral (\ref{borel}) is \cite{hardy,carl} an analytic function of $g$  
in the cut $g$ plane: $|{\rm arg}(g)|<\pi$. So a dispersion relation (using
the discontinuity across the cut along the negative $g$ axis) can be used to  
{\it define} the imaginary part of $f(g)$ for negative values of the expansion  
parameter:
\begin{eqnarray}
{\rm Im} f(-g)\sim\frac{\pi}{\rho g}\exp[-\frac{1}{\rho g}]
\label{imag}
\end{eqnarray}
The imaginary contribution (\ref{imag}) is non-perturbative (it clearly does  
not have an expansion in positive powers of $g$) and has important physical
consequences. Note that (\ref{imag}) is consistent with a principal parts
prescription for the pole that appears on the $s>0$ axis if we make the formal  
manipulations as in (\ref{borel}):
\begin{eqnarray}
\sum_{n=0}^\infty \rho^n n! \, g^n \sim \frac{1}{\rho g}\int_0^\infty
ds\,\left(\frac{1}{1-s}\right)\exp\left[-\frac{s}{\rho g}\right]
\label{non}
\end{eqnarray}

Similar formal arguments can be applied to the case when the expansion
coefficients have leading behaviour (\ref{general}). Then the leading Borel
approximation is
\begin{eqnarray}
f(g)\sim \frac{1}{\mu}\, \int_0^\infty \frac{ds}{s} \, \left(\frac{1}{1+s}\right)
\left(\frac{s}{\rho g}\right)^{\nu/\mu}\,
\exp\left[-\left(\frac{s}{\rho g}\right)^{1/\mu}\right]
\label{genborel}
\end{eqnarray}
For the corresponding non-alternating case, when $g$ is negative, the leading  
imaginary contribution is
\begin{eqnarray}
{\rm Im} f(-g)\sim\frac{\pi}{\mu}\left(\frac{1}{\rho g} \right)^{\nu/\mu}
\exp\left[-\left(\frac{1}{\rho g}\right)^{1/\mu}\right]
\label{genimag}
\end{eqnarray}
Note the separate meanings of the parameters $\rho$, $\mu$ and $\nu$ that
appear in the formula (\ref{general}) for the leading large-order growth of
the expansion coefficients. The constant $\rho$ clearly combines with $g$ as  
an effective expansion parameter. The power of the exponent in
(\ref{genimag}) is determined by $\mu$, while the power of the prefactor in
(\ref{genimag}) is determined by the ratio $\frac{\nu}{\mu}$.

It must be stressed that these formulas (\ref{genborel}) and (\ref{genimag})  
are formal, being based on assumed analyticity properties of the function
$f(g)$. The Borel dispersion relations could be complicated by the appearance  
of additional poles and/or cuts in the complex $g$ plane, signalling new
physics \cite{thooft}. In certain special cases these analyticity assumptions  
can be tested rigorously, but we have in mind the situation in which one is
confronted with the expansion coefficients $a_n$ of a perturbative expansion,  
without corresponding information about the function that this series is
supposed to represent. This is a common circumstance in physical applications  
of perturbation theory. For example, Borel techniques have recently been used  
to study the divergence of the derivative expansion for QED effective actions  
in inhomogeneous backgrounds \cite{dh}.

Returning to the Euler-Heisenberg effective Lagrangian, the question of
whether the perturbative expansion is alternating or non-alternating is
directly relevant. For a uniform magnetic background, the one-loop
Euler-Heisenberg series (\ref{eh}) is precisely of the form (\ref{exp}) with  
$g=(\frac{eB}{m^2})^2$. Moreover, from (\ref{growth}) the expansion
coefficients $a_n^{(1)}$ have leading large-order behaviour of the form
(\ref{general}), with $\rho=\frac{1}{\pi^2}$, and $\mu=\nu=2$. In fact, taking  
into account the sub-leading corrections indicated in (\ref{growth}), the
proper-time integral representation (\ref{proper}) is precisely the Borel sum,  
using (\ref{genborel}), of the divergent series (\ref{eh}) \cite{dh}. For a
uniform {\it electric} background, the only difference perturbatively is that  
$B^2$ is replaced by $-E^2$; that is, $g=(\frac{eB}{m^2})^2$ is replaced by
$-g=-(\frac{eE}{m^2})^2$. So the perturbative one-loop Euler-Heisenberg
series (\ref{eh}) becomes non-alternating. Then from (\ref{genimag}), with
$\rho=\frac{1}{\pi^2}$ and $\mu=\nu=2$, we immediately deduce the leading
behaviour of the imaginary part of the one-loop Euler-Heisenberg effective
Lagrangian:
\begin{eqnarray}
{\rm Im} {\cal L}^{(1)} \sim
\frac{m^4}{8\pi^3}\left(\frac{eE}{m^2}\right)^2\,
\exp\left[-\frac{m^2\pi}{eE}\right]
\label{leadingimag}
\end{eqnarray}
This imaginary part has direct physical significance - it gives half the
electron-positron pair production rate in the uniform electric field $E$
\cite{schwinger,nikishov,greiner}.
Actually, since we also know the
sub-leading corrections (\ref{growth}) to the leading large-order behaviour of  
the expansion coefficients $a_n^{(1)}$, we can apply (\ref{genimag})
successively to go beyond the leading behaviour in (\ref{leadingimag}):
\begin{eqnarray}
{\rm Im} {\cal L}^{(1)} \sim  \frac{m^4}{8\pi^3}
\left(\frac{eE}{m^2}\right)^2\, \sum_{k=1}^\infty \frac{1}{k^2}
\,\exp\left[-\frac{m^2\pi k}{eE}\right]
\label{fullimag}
\end{eqnarray}
This is Schwinger's classic result \cite{schwinger} for the imaginary part of  
the one-loop effective Lagrangian in a uniform electric field $E$.
To elucidate the physical meaning of the individual terms of this
series it is useful to employ the
following alternative
representation due to Nikishov \cite{nikishov},

\begin{eqnarray}
 {2\over \hbar} {\rm Im} {\cal L}^{(1)}
VT &=&
-\sum_r \int {d^3p V\over {(2\pi\hbar)}^3}
{\rm ln}(1-\bar n_p),\nonumber\\
\bar n_p &=& \exp \Bigl(-\pi {m^2 + p_{\perp}^2\over eE}\Bigr)
\label{repnikishov}
\end{eqnarray}
Here $\bar n_P$ is the mean number of pairs produced by the field
in the state with given momentum $p$ and spin projection $r$.
An expansion of the logarithm in $\bar n_P$ and term-by-term
integration leads back to Schwinger's
formula (\ref{fullimag}). Thus the leading term
in this formula can be interpreted as the mean number $\bar n_P$
of pairs in the unit 4-volume $VT$, while the higher ($k\ge 2$)
terms describe the coherent creation of $k$ pairs.

Pair creation can occur for any value of the electric field
strength, though due to the exponential suppression factors
one is presently still far away from being able to observe
spontaneous pair creation by macroscopic fields in the laboratory.
However, it can be arranged for
electrons traversing the focus of a terawatt laser to see a
critical field in their rest frame.
This has recently led to the first observation of
pair creation in a process involving only real photons
\cite{burkeetal}.

For the one-loop Euler-Heisenberg QED effective Lagrangian, this large-order  
perturbation theory analysis is greatly simplified by the fact that we know
the exact formula (\ref{growth}) for the expansion coefficients $a_n^{(1)}$.  
This will not be the case below, when we discuss the two-loop Euler-Heisenberg  
effective Lagrangian. So, for the sake of numerical comparison, we compare
the exact one-loop coefficients $a_n^{(1)}$ with their leading large-order  
behaviour.
The coefficients are listed in Table 1 up to $n=15$. Since the growth is fast,  
it is convenient to compare the logarithms, as is done in Figure 1. With 16  
terms it is
straightforward to fit the the values of $\rho$, $\mu$ and $\nu$ appearing in  
(\ref{general}); moreover, there is sufficient accuracy to fit the overall
coefficient $\frac{1}{8 \pi^4}$. In Figure 1 we plot $A_n^{(1)} \equiv \log
|a_n^{(1)}|$, and $C_n^{(1)}=\log[ \Gamma(2n+2)/(8\pi^{2n+4})]$. The agreement  
is spectacular, already for $n=0$. Indeed, on this scale the two plots are
indistinguishable. To go beyond the leading large-order behaviour, we plot the  
difference $A_n^{(1)}-C_n^{(1)}$. This can be fitted to the correct form
$\log(1+\frac{1}{2^{2n+4}})\approx \frac{1}{2^{2n+4}}$, with remarkable
accuracy for $n\geq 2$, as illustrated in Figure 2.


\begin{table*}[p]
\caption[t1]{The one-loop Euler-Heisenberg coefficients $a_n^{(1)}$ from
(\protect{\ref{growth}}) and their magnitudes $|a_n^{(1)}|$. The last two
columns list the calculated two-loop Euler-Heisenberg coefficients $a_n^{(2)}$  
in (\protect{\ref{2leh}}), and their magnitudes $|a_n^{(2)}|$. Note that
both $a_n^{(1)}$ and $a_n^{(2)}$ alternate in sign and grow factorially in
magnitude.}
\vspace{20pt}
\begin{tabular}{c|cccc}
$n$ & $a_n^{(1)}$ & $|a_n^{(1)}|$ & $a_n^{(2)}$ & $|a_n^{(2)}|$ \\ \tableline
0  & $\frac{1}{720}$ & 0.00138889 & $\frac{64}{81}$ & 0.790123   \\ \\
1  & $-\frac{1}{1260}$  &   0.000793651 & $-\frac{1219}{2025}$  & 0.601975\\ \\
2  & $\frac{1}{630}$  &   0.0015873 & $\frac{135308}{99225}$  &   1.36365\\\\
3  & $-\frac{2}{297}$ & 0.00673401 & $-\frac{791384}{127575}$ & 6.20328\\\\
4  & $\frac{11056}{225225}$  &   0.0490887 & $\frac{8519287552}{180093375}$   
&   47.3048 \\\\
5  & $-\frac{64}{117}$  &   0.547009 & $-\frac{167313653536}{307432125}$  &    
544.23 \\\\
6  & $\frac{231488}{26775}$ & 8.64568 & $\frac{132482118784}{15035625}$ &
8811.21 \\\\
7  & $-\frac{11229952}{61047}$  &   183.956 &
$-\frac{21154062270001664}{110718132525}$  &   191062 \\\\
8  & $\frac{715206656}{141075}$  &   5069.69 &
$\frac{3457277998817366503424}{646561329789375}$  &  5.34718~$10^6$\\\\
9  & $-\frac{1272758272}{7245}$ & 175674 &
$-\frac{6386544394068463796224}{34029543673125}$ & 1.87676~$10^8$ \\\\
10  & $\frac{7745178533888}{1036035}$  &   $7.47579$~$10^6$ &
$\frac{1889413511360959841329676288}{234021171840080625}$  &
$8.07369$~$10^9$ \\\\
11  & $-\frac{86236332032}{225}$  &   $3.83273$~$10^8$ &
$-\frac{175667006973780978406129664}{420447667696875}$  &
$4.17809$~$10^{11}$ \\\\
12  & $\frac{3557587835420672}{152685}$ & $2.33002$~$10^{10}$ &
$\frac{298448176892800474966309470208}{11654809348557375}$ &
$2.56073$$10^{13}$ \\\\
13  & $-\frac{7227491505462575104}{4361049}$  & $1.65728$~$10^{12}$ &
$-\frac{26169081938037425304875653922816}{14265121494999375}$ &
$1.83448$~$10^{15}$ \\\\
14  & $\frac{16167618040230313984}{118575}$  &   $1.36349$~$10^{14}$ &
$\frac{3428377058882578209519972928482967552}{22571190592506430875}$  &
1.51892~$10^{17}$   \\\\
15  & $-\frac{1271953705305899008}{99}$ & $1.2848$~$10^{16}$ &
$-\frac{401755841159672623527223409633644249088}{27910612022991823125}$ &
$1.43944$~$10^{19}$ \\
\end{tabular}
\label{t1}
\end{table*}

\pagebreak

\begin{figure}[htb]
\vskip -1cm
\centering{\epsfig{file=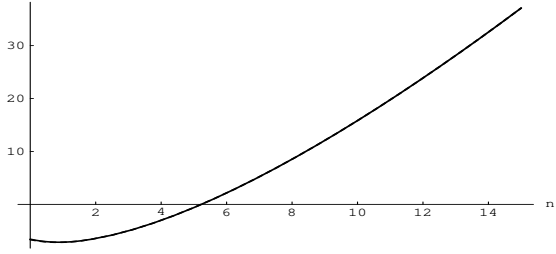, width = 5in, height=5in}}
\vskip -3in
\caption{Plots of $A_n^{(1)}=\log(|a_n^{(1)}|)$ and $C_n^{(1)}=\log[2
\Gamma(2n+2)/(8 \pi^{2n+4})]$ up to $n=15$. The two plots are
indistinguishable on this scale.}
\label{f1}
\end{figure}
\begin{figure}[htb]
\vskip -1cm
\centering{\epsfig{file=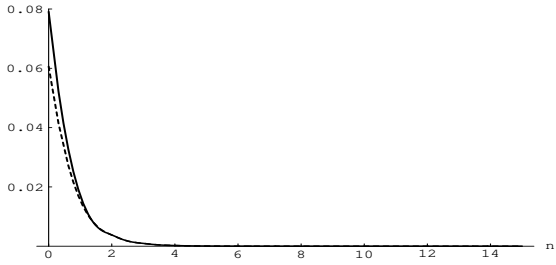, width = 5in, height=5in}}
\vskip -3in
\caption{Plot (solid line) of the difference $A_n^{(1)}-C_n^{(1)}$ between
$A_n^{(1)}=\log(|a_n^{(1)}|)$ and the leading large order fit $C_n^{(1)}=\log[  
\Gamma(2n+2)/(8\pi^{2n+4})]$ from (\protect{\ref{growth}}), up to $n=15$. The  
dashed line shows the fit of this difference to the function
$\log(1+\frac{1}{2^{2n+4}})$, which is excellent already for $n\geq 2$.}
\label{f2}
\end{figure}

\section{The Two-Loop Euler-Heisenberg Effective Lagrangian}

We now turn to the two-loop Euler-Heisenberg effective
Lagrangian, describing the effect of
a single photon exchange in the electron loop.
This quantity was first studied
by Ritus \cite{ritus}. Using the exact electron propagator
in a constant field found by Fock \cite{fock} and
Schwinger \cite{schwinger}, and a proper-time cutoff as
the UV regulator, he obtained the on-shell renormalized
${\cal L}^{(2)}$ in terms
of a certain two-parameter integral. From this integral
the imaginary part of the Lagrangian was then extracted by
a painstaking analysis of its analyticity properties,
yielding a representation analogous to Schwinger's one-loop
formula (\ref{fullimag})
\cite{lebrit}. Adding up the one-loop and the
two-loop contributions to the imaginary part the result
reads

\begin{eqnarray}
{\rm Im} {\cal L}^{(1)} +
{\rm Im} {\cal L}^{(2)} &\sim&  \frac{m^4}{8\pi^3}
\beta^2\,
\sum_{k=1}^\infty
\Bigl[
\frac{1}{k^2}
+\alpha\pi K_k(\beta)
\Bigr]
\,\exp\left[-\frac{\pi k}{\beta}\right]
\label{fullimag2loop}
\end{eqnarray}
where $\beta = {eE\over m^2}$. For the function $K_k(\beta)$
the following small $\beta$ - expansion was obtained
in \cite{lebrit},

\begin{eqnarray}
K_k(\beta) &=& -{c_k\over \sqrt{\beta}} + 1 + {\rm O}(\sqrt{\beta})
 \nonumber\\
c_1 = 0,\quad && \quad
c_k = {1\over 2\sqrt{k}}
\sum_{l=1}^{k-1} {1\over \sqrt{l(k-l)}},
\quad k \geq 2
\nonumber\\
\label{expK}
\end{eqnarray}
According to \cite{ritus} the physical interpretation
of the individual terms in the series in terms of coherent
multipair creation can be carried over to the two-loop
level. This requires one to absorb the term linear in
$\alpha$ into the exponential factor, rewriting

\begin{eqnarray}
\Bigl[
\frac{1}{k^2}
+\alpha\pi K_k\bigl({eE\over m^2}\bigr)
\Bigr]
\,\exp\left[-{k\pi m^2\over eE}\right]
\sim
\frac{1}{k^2}
\exp\left[-{k\pi m^2_{\ast}\over eE}\right]
\label{shiftm}
\end{eqnarray}
The individual terms in the expansion
(\ref{fullimag2loop}) of $K_k(\beta)$ are then to be
absorbed into the mass shift $m_{\ast}-m$. For the lowest
order terms in this expansion, those given
in (\ref{fullimag2loop}),
the physical meaning of the corresponding
mass shifts in terms of the coherent pair production
picture is discussed in \cite{ritus}.
For example, the leading $``1''$ in the expansion of
$K_1(\beta)$ after exponentiation yields a mass shift
that can be identified as the {\sl classical}
change in mass caused for one particle in a created pair
by the acceleration due to its partner.

Assuming this exponentiation to work one can, of course,
obtain some partial information on the higher-loop corrections
to the imaginary part. More remarkably, since the above
physical interpretation requires the mass appearing in the
exponent to be the physical one, it allows
one to determine the physical renormalized mass from
an inspection of the renormalized Lagrange function
alone, rather than by a calculation of the (lower
order) electron self energy.

Following the pioneering work by Ritus and his collaborators,
a first recalculation of the Euler-Heisenberg Lagrangian
was done by Dittrich and Reuter \cite{dittrich} for the magnetic
field case.
The more recent recalculation in \cite{rss}
showed that the two previous results were
actually incompatible, and differed precisely by a
finite electron mass renormalisation.
All three calculations had been
done using a proper-time cutoff rather than dimensional
regularisation. This cutoff leads to relatively simple integrals,
but due to its non-universality
makes it, at the two-loop level,
already non-trivial to determine the physical renormalized
electron mass.
A fourth calculation of this quantity
\cite{frss}, now using dimensional regularisation,
yielded complete agreement with Ritus's result
after a perturbative expansion of both results
in powers of the $B$ field had been performed.
In the following
we will push the same calculation to ${\rm
O}(B^{34})$,
and analyze the rate of growth of the expansion coefficients.

\section{Borel Analysis of the Two-Loop Euler-Heisenberg Effective Lagrangian}

The world-line expression for the two-loop on-shell renormalized
Euler-Heisenberg effective Lagrangian is \cite{rss,frss}:
\begin{eqnarray}
{\cal L}^{(2)}&=& \alpha\, \frac{m^4}{(4\pi)^3} \left(\frac{eB}{m^2}\right)^2  
\int_0^\infty \frac{ds}{s^3}\, e^{-m^2 s/(eB)}\, \int_0^1 du\, \left[
L(s,u)-2s^2+\frac{6}{u(1-u)}\left(\frac{s^2}{{\rm sinh}^2s} +s\, {\rm
coth}s\right)\right] \nonumber\\
&&\qquad -12\alpha\, \frac{m^4}{(4\pi)^3} \frac{eB}{m^2} \int_0^\infty
\frac{ds}{s}\, e^{-m^2 s/(eB)}\, \left[ {\rm
coth}s-\frac{1}{s}-\frac{s}{3}\right]\, \left[
\frac{3}{2}-\gamma-\log\left(\frac{m^2 s}{eB}\right) +\frac{eB}{m^2 s}\right]
\label{2l}
\end{eqnarray}
Here $\alpha\approx \frac{1}{137}$ is the fine-structure constant, and
$\gamma=0.5772...$ is Euler's constant.
The function $L(s,u)$ appearing in the integrand of the first term in
(\ref{2l}) is defined by the following relations:

\begin{eqnarray}
L(s,u)&=&s\,{\rm coth}s\left[ {\log\left(\frac{u(1-u)}{G(u,s)}\right)
\over [u(1-u)-G(u,s)]^2}\, F_1 +{F_2\over G(u,s)[u(1-u)-G(u,s)]} +
{F_3\over u(1-u)[u(1-u)-G(u,s)]}\right]\nonumber\\
G(u,s)&=&{{\rm cosh} s- {\rm cosh}((1-2u)s)\over 2 s\, {\rm sinh}s}\nonumber\\
F_1&=&4 s({\rm coth}s-{\rm tanh}s) G(u,s) - 4u(1-u)\nonumber\\
F_2&=&2(1-2u) \frac{{\rm sinh}((1-2u)s)}{{\rm sinh}s}
+s (8 {\rm tanh}s-4 {\rm coth} s)G(u,s)-2\nonumber\\
F_3&=&4u(1-u)-2(1-2u)\frac{{\rm sinh}((1-2u)s)}{{\rm sinh}s}
-4 s \, G(u,s) {\rm tanh} s+2
\label{defs}
\end{eqnarray}

The second term in the two-loop expression (\ref{2l}) is generated by
the one-loop electron mass renormalisation, which at the two-loop
level becomes necessary in addition to the photon wave function renormalisation.
For this mass renormalization term we have found the following exact expansion:
\begin{eqnarray}
&&\frac{eB}{m^2} \int_0^\infty \frac{ds}{s}\, e^{-m^2 s/(eB)}\, \left[ {\rm
coth}s-\frac{1}{s}-\frac{s}{3}\right]\, \left[
\frac{3}{2}-\gamma-\log\left(\frac{m^2 s}{eB}\right) +\frac{eB}{m^2 s}\right]
 \nonumber\\
&&\qquad\qquad = \left(\frac{eB}{m^2}\right)^4 \sum_{n=0}^\infty {2^{2n+4}
{\cal B}_{2n+4}\over (2n+4)(2n+3)} \left(\frac{3}{2}-\gamma -\psi(2n+2)\right)  
\, \left(\frac{eB}{m^2}\right)^{2n}
\label{2lmass}
\end{eqnarray}

Here ${\cal B}_{2n}$ are the Bernoulli numbers, and
$\psi(x)=\frac{\Gamma^\prime(x)}{\Gamma(x)}$ is the digamma function
\cite{gradshteyn}. We have not succeeded in finding a closed-form
expression for the expansion coefficients arising from the expansion
of the double integral in (\ref{2l}), although we suspect that one may
exist. Instead, we have made an algebraic expansion of this integral,
using MATHEMATICA and MAPLE.
When combined with the exact expansion (\ref{2lmass}) of the mass
renormalization term we obtain an
expansion of the form:
\begin{eqnarray}
{\cal L}^{(2)} = \alpha\, \frac{m^4}{(4\pi)^3} \left(\frac{eB}{m^2}\right)^4  
\sum_{n=0}^\infty a_n^{(2)} \, \left(\frac{eB}{m^2}\right)^{2n}
\label{2leh}
\end{eqnarray}
The expansion coefficients $a_n^{(2)}$ are listed in Table 1, up to $n=15$.
Note that those coefficients are in some sense less universal than their
one-loop counterparts, since they depend on the one-loop
normalization condition imposed on the renormalized electron mass.

Several comments are in order. First, the two-loop expansion coefficients
$a_n^{(2)}$ alternate in sign, just as in the one-loop magnetic background
case (\ref{growth}). Second, the magnitude $|a_n^{(2)}|$ is clearly growing
factorially fast with $n$. Thus, the two-loop Euler-Heisenberg series
(\ref{2leh}) is a divergent series,
as is the one-loop Euler-Heisenberg series
(\ref{eh}).
Note also that for each series the smallest magnitude coefficient is reached  
already for $n=1$, after which the coefficients begin to increase rapidly in  
magnitude.

To extract the leading large-n growth of $|a_n^{(2)}|$ we fit $a_n^{(2)}$ to  
the form in (\ref{general}). Once again, it is convenient to work with the
logarithm $D_n=\log(|a_n|)$ since the growth is so rapid. It is relatively
straightforward to find that $\mu=\nu=2$ and $\rho=\frac{1}{\pi^2}$. It is
more difficult to fit the overall coefficient, but if we assume this is a
simple power of $\pi$ then our best fit for the leading large-order growth of  
the two-loop expansion coefficients in (\ref{2leh}) is:
\begin{eqnarray}
a_n^{(2)}\sim (-1)^n \frac{16}{\pi^2} \, \frac{\Gamma(2n+2)}{\pi^{2n}}
\label{2lgrowth}
\end{eqnarray}
This leading fit is displayed in Figure 3, in terms of
$A_n^{(2)}\equiv\log(|a_n^{(2)}|)$ . The fit is not as good as the one-loop
fit shown in Figure 1, but it is still very good.

Note the remarkable similarity of the leading large-order growth
(\ref{2lgrowth}) of the two-loop expansion coefficients to the leading
large-order growth of the one-loop expansion coefficients in (\ref{growth}).  
The only difference is the overall coefficient. The parameters $\rho$, $\mu$  
and $\nu$ in the general form (\ref{general}) are identical. Using the Borel  
technique to relate this leading growth rate to the leading non-perturbative  
imaginary part of the effective Lagrangian in a uniform electric field $E$, we  
deduce that the two-loop leading imaginary part is proportional to the
one-loop leading imaginary part (\ref{leadingimag}). In fact, from
(\ref{genimag}) and (\ref{2lgrowth}), we find the leading contribution
\begin{eqnarray}
{\rm Im} {\cal L}^{(2)} \sim  \alpha\, \pi\,
\frac{m^4}{8\pi^3}\left(\frac{eE}{m^2}\right)^2\,
\exp\left[-\frac{m^2\pi}{eE}\right]
\label{2lleading}
\end{eqnarray}

\begin{figure}[htb]
\vskip -1cm
\centering{\epsfig{file=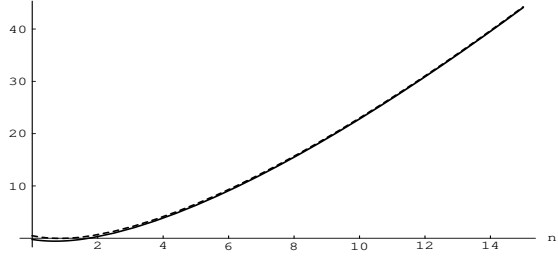, width = 5in, height=5in}}
\vskip -3 in
\caption{Plots of $A_n^{(2)}=\log(|a_n^{(2)}|)$ (solid line), and
$C_n^{(2)}=\log[16 \Gamma(2n+2)/\pi^{2n+2}]$ (dashed line), up to $n=15$.
While the agreement is not as good as for the one-loop case shown in Figure 1,  
the fit is still very good, even at low orders in $n$.}
\label{f3}
\end{figure}

This has exactly the same dependence on the electric field $E$ as the
one-loop case. So to two-loop order the leading non-perturbative behaviour of  
the imaginary part of the effective Lagrangian is:
\begin{eqnarray}
{\rm Im} \left({\cal L}^{(1)}+{\cal L}^{(2)}\right) \sim
\left(1+\alpha\,\pi\right)\, \frac{m^4}{8\pi^3}\left(\frac{eE}{m^2}\right)^2\,  
 \exp\left[-\frac{m^2\pi}{eE}\right]
\label{leadingsum}
\end{eqnarray}
This agrees with the leading term of Ritus's formula (\ref{fullimag2loop}).

\begin{figure}[htb]
\vskip -1cm
\centering{\epsfig{file=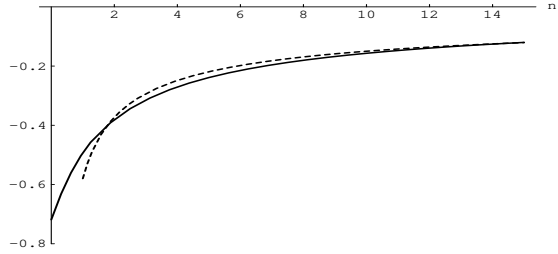, width = 5in, height=5in}}
\vskip -3in
\caption{Plot (solid line) of the difference $A_n^{(2)}-C_n^{(2)}$ between
$A_n^{(2)}=\log(|a_n^{(2)}|)$ and the leading large order fit from
(\protect{\ref{2lgrowth}}), $C_n^{(2)}=\log[16 \Gamma(2n+2)/\pi^{2n+2}]$,
up to $n=15$. The dashed line represents the fit in (\protect{\ref{2lcorr}}).}
\label{f4}
\end{figure}

To go beyond this leading term we need to look at corrections to the leading  
behaviour in (\ref{2lleading}). In Figure 4 we plot the difference of the
logarithms, and we see that the $n$ dependence is much more gentle than the
rapid fall-off found in the one-loop case, which was plotted in Figure 2. In  
fact, from the terms up to $n=15$, we obtain the fit
\begin{eqnarray}
a_n^{(2)}\sim (-1)^n \frac{16}{\pi^2} \, \frac{\Gamma(2n+2)}{\pi^{2n}}\left[  
1-\frac{0.44}{\sqrt{n}}+\dots \right]
\label{2lcorr}
\end{eqnarray}
This is a considerably weaker $n$ dependence than is found for the first
correction in the one-loop case (\ref{growth}). This means that in the
two-loop case the dominant corrections are to the prefactor in the leading
behaviour (\ref{2lleading}). This is in contrast to the one-loop case
(\ref{fullimag}) where the first correction to the leading behaviour is
exponentially suppressed. Indeed, applying the Borel relations, the correction  
term (\ref{2lcorr}) leads to
\begin{eqnarray}
{\rm Im} \left({\cal L}^{(1)}+{\cal L}^{(2)}\right) \sim
\left(1+\alpha\,\pi\left[1- (0.44) \sqrt{\frac{2eE}{\pi
m^2}}+\dots\right]\right)\, \frac{m^4}{8\pi^3}\left(\frac{eE}{m^2}\right)^2\,   
\exp\left[-\frac{m^2\pi}{eE}\right]
\label{nextsum}
\end{eqnarray}
We emphasize that the fit in (\ref{2lcorr}) is based on a simple fit
to the first 16 two-loop coefficients. Nevertheless, the structure of
(\ref{nextsum}) conforms already to the form of Ritus's expansion in  
eq.(\ref{fullimag2loop}). It would be
interesting to probe this correction term in more detail by a further
study of the analyticity properties of the
integral representations \cite{ritus,rss,frss}
of the two-loop Euler-Heisenberg effective Lagrangian, or by looking at still  
higher orders in perturbation theory.

\section{Concluding Remarks}

Our analysis also permits us to study the dependence of (\ref{2lleading}), the  
leading non-perturbative imaginary contribution to the effective Lagrangian,   
on the electron mass renormalisation. In the world-line expression (\ref{2l})  
for the two-loop Euler-Heisenberg effective Lagrangian, a finite change of
the renormalised electron mass would amount to an arbitrary change
of the constant ${3\over 2} -\gamma$ appearing in the second
bracket of the second term. For example, in \cite{rss} it had been shown
that the renormalised Lagrangian obtained by \cite{dittrich} differs
from (\ref{2l}) precisely by a replacement of ${3\over 2}$
by ${5\over 6}$. A separate study of the contributions of the first
and second term in (\ref{2l}) shows that, due to cancellations
between both terms, the leading large-n growth
of their sum is smaller than for each term separately.
However this property holds true only if the renormalised electron
mass is the physical one.

\begin{figure}[htb]
\vskip -1cm
\centering{\epsfig{file=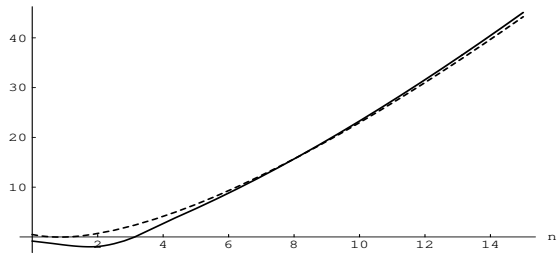, width = 5in, height=5in}}
\vskip -3in
\caption{Plot (dashed line) of the correct leading behaviour,
$C_n^{(2)}=\log[16 \Gamma(2n+2)/\pi^{2n+2}]$, of the logarithm of the
magnitude of the two-loop expansion coefficients; as compared to (solid line)  
$\tilde{A}_n^{(2)}=\log(|\tilde{a}_n^{(2)}|)$, obtained from the two-loop
coefficients $\tilde{a}_n^{(2)}$ extracted from the result of 
\protect{\cite{dittrich}}. One can clearly see the
divergence between these two curves at large $n$, indicating
a difference in
the leading growth of the coefficients.}
\label{f5}
\end{figure}

For definiteness, in Figure 5 we compare the
correct leading growth (\ref{2lgrowth}) of the two-loop coefficients with the  
coefficients obtained by
expanding out the two-loop result of \cite{dittrich}.
The difference in the leading large-n growths is obvious.
Thus, in agreement with the analysis of \cite{ritus}, we find that
if and only if expressed in terms of the true electron
mass will the imaginary part of the renormalised two-loop Lagrangian show
the same exponential suppression factor $\exp[-\pi m^2/(eE)]$ as for the
one-loop Lagrangian.

To summarize, we have constructed the imaginary part of the
two-loop QED Euler-Heisenberg Lagrangian in a constant electric
field by a computer based calculation of its weak
field expansion together with Borel summation techniques.
The knowledge of the first 16 coefficients has turned out to be
sufficient to verify the structure of the leading $(k=1)$
term in Ritus's eq.(\ref{fullimag2loop}), and to obtain a numerical value
for the first $O(\sqrt{eE/m^2})$ correction contained
in that formula. The method used here is significantly
simpler than the one in \cite{ritus,lebrit}, where the imaginary
part was obtained by an analysis of the analyticity
properties of the two-loop parameter integrals.

In particular, we have seen that the large order behaviour of the two-loop
coefficients is the same (up to an overall constant factor) as the one-loop
case. This means that the leading contribution to the imaginary part of the
two-loop effective Lagrangian has the same form as in the one-loop
case.
This gives a new perspective to Ritus's
arguments \cite{ritus}
that the true renormalized electron mass $m$ is such that the
leading exponential factor in the pair production rate is
$\exp[-\pi m^2/(eE)]$. Since those arguments pertain to arbitrary loop orders,
and the leading exponential factor is directly related
to the leading growth rate in the weak field expansion,
they also lead one to expect that the Euler-Heisenberg Lagrangian
may be amenable to this type of Borel analysis at any fixed loop order in  
perturbation
theory.

\vskip 1cm

\section{Acknowledgements:}
C. S. would like to thank R. Stora for discussions.
This work has been supported in part (GD) by the U.S. Department of Energy
grant DE-FG02-92ER40716.00.

\end{document}